\documentclass[twocolumn,showpacs,showkeys,preprintnumbers,
               amsmath,amssymb,floatfix,10pt]{revtex4}
\usepackage{graphicx}
\usepackage{dcolumn}
\usepackage{bm}
\usepackage{graphicx}
\usepackage{subfigure}
\usepackage{amsmath}
\usepackage{amssymb}
\usepackage{wrapfig}
\usepackage{color}
\usepackage[english]{babel}
\usepackage[warning,math]{easyeqn}
\usepackage[thinlines]{easybmat}

\def\<{\langle}
\def\>{\rangle}
\begin{document}
%
\preprint{}
\title{Simplified flexibility analysis of proteins}
\author{Yves-Henri Sanejouand}
\email{yves-henri.sanejouand@univ-nantes.fr}
\affiliation{UFIP,
FRE-CNRS 3478, Facult\'e des Sciences et des Techniques,
Nantes, France.}
%
%
%
%
%
%
\maketitle
%
\section*{Abstract} 
A simple way to get insights about the possible functional motions
of a protein is to perform a normal mode analysis (NMA). 
Indeed, it has been shown that low-frequency modes
thus obtained are often closely related to domain motions
involved in protein function.
Moreover, because protein low-frequency modes are known to be robust, 
NMA can be performed using 
coarse-grained models. As a consequence, it can be done for large ensembles
of conformations as well as for large systems, like the ribosome, whole virus capsids, \textit{etc}. 
Unexpectedly, on the high-frequency side, modes obtained with cutoff-based coarse-grained models
also seem able to provide useful insights on protein dynamical properties. 
\section{Introduction}

In order to understand the function of a protein, the knowledge of its structure
is of utmost importance. However, it is becoming more and more clear that in most cases this
is not enough and that the relevant information needed is its {\it structural ensemble},
that is, a fair sample of the set of conformations a protein can visit
at room temperature.

A straightforward way to obtain this information is to perform a {\it long enough} molecular dynamics (MD)
simulation in explicit solvent. In practice, noteworthy in the case of enzymes,
the timescale that has to be reached is at least of the order of the microsecond
since the most efficient enzymes, like catalase, have turn-overs of this order of magnitude.
Nowadays, for standard size proteins, MD simulations that long are routinely performed
on supercomputers. Actually, using a dedicated ASICS-based supercomputer, the millisecond timescale has recently been reached, 
in the case of the small BPTI model system~\cite{Shaw:10}.
Moreover, starting from random initial configurations, accurate {\it ab initio} folding of several 
fast-folding peptides has been obtained~\cite{Shaw:10,Shaw:11},
revealing all the potential of the brute force approach for the years to come.

However, even with a supercomputer, obtaining the relevant structural ensemble through
this approach often takes months. Moreover, this can hardly be done for very large
systems, like the ribosome, whole virus capsids, \textit{etc}. So, simplified methods
are still welcome, noteworthy because, as a consequence of their low computational cost,
they can be implemented on web-servers, where anybody can give them a try.

\section{Normal Mode Analysis}

\subsection{Background}

A simple way to study the flexibility of a molecule
is to perform a normal mode analysis (NMA). This method rests upon
the fact that the classical equations of motion for a set of $N$ atoms can be solved analytically 
when the displacements of the atoms in 
the vicinity of their equilibrium positions are small enough.
As a consequence,
$V$, the potential energy of the studied system, can be approximated by the first terms of a Taylor series.
Moreover, because the system is expected to be at equilibrium~\footnote{Or on a saddle point.}, the term
with the first derivatives of the energy (the forces) can be dropped, so that:
\begin{equation}
V = \frac{1}{2} \sum_{i=1}^{3N} \sum_{j=1}^{3N} \left(\frac{\partial^2 V}
  {\partial r_i \partial r_j}\right)_0 (r_i - r_i^0) (r_j - r_j^0) 
\label{eq:approx}
\end{equation}
where $r_i$ is the $i^{th}$ coordinate of the system, $r_i^0$ being its equilibrium value.

In other words, within the frame of this approximation,
the potential energy of a system can be written as a quadratic form.
In such a case, it is quite straightforward to show that the
equations of atomic motion have the following solutions~\cite{Goldstein:50,Wilson:55}:
\begin{equation}
r_i(t) = r_i^0 + \frac{1}{\sqrt{m_i}}
       \sum_{k=1}^{3N} C_k a_{ik} cos( 2 \pi \nu_k t + \Phi_k )
\label{eq:fluct}
\end{equation}
where $m_i$ is the atomic mass and where $C_k$ and $\Phi_k$, the amplitude and phasis of the so-called normal mode
of vibration $k$, depend upon the initial conditions, that is, upon atomic positions and velocities at $t=0$.

In practice, $\nu_k$, the frequency of mode $k$, is obtained through the 
$k^{th}$ eigenvalue of the mass-weighted Hessian of the potential energy,
that is, the matrix whose elements are the mass-weighted second derivatives of the energy,
the $a_{ik}$'s being the coordinates of the corresponding eigenvector.

\begin{figure}[t!]
\includegraphics[width=8.5 truecm,clip]{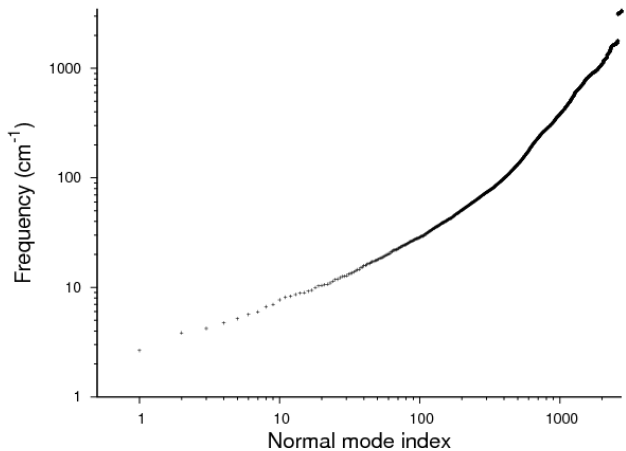}
\caption{\small \label{Fig:spectre}
Spectrum of the HIV protease, as obtained
using standard normal mode analysis. The structure considered is the monomer
found in PDB 1HHP, after energy-minimization. High-frequency modes are localized motions
of chemically bonded atoms. The highest-frequency ones correspond to motions
in which a hydrogen atom is involved. There are only a few (164) of them 
because an extended-atom model was considered~\cite{EEF1}.
}
\end{figure}

In Fig.~\ref{Fig:spectre}, the frequency spectrum thus obtained for the protease of human immunodeficiency virus (HIV)
is shown, when a standard empirical energy function is used, namely, as available in CHARMM~\cite{Charmm}. 
The modes with the highest frequencies 
correspond to motions of pairs of atoms that are chemically bonded
and the reason why these frequencies are much higher than others is because
a hydrogen (\textit{i.e.}, a light particle) is involved in these bonds.
Generally speaking, modes with high frequencies are localized, that is,
only a few atoms are moving significantly at these frequencies.   

On the contrary, on the low-frequency part of the spectrum, modes tend to involve 
large parts of the structure. This is probably why their comparison with
protein functional motions was undertaken, the latter being often described as relative motions of structural domains~\cite{Gerstein:98}. 
As a matter of fact, in the first NMA study of a protein with well defined domains,
namely, the human lysozyme, the lowest-frequency mode ($\nu_1$ = 3.6 cm$^{-1}$)
was found to correspond to a hinge-bending motion~\cite{Brooks:85}.
Later on, the lowest frequency modes of citrate synthase~\cite{Marques:95} 
and haemoglobin~\cite{Guilbert:95} were indeed found able to provide a fair description
of the rather complex and large-amplitude motion these proteins experience upon ligand binding.

\subsection{Technical issue}

Getting the eigenvalues and eigenvectors of a large matrix can prove challenging,
in particular because methods available in mathematical libraries usually require
the storage of the whole matrix in the computer memory. For a small protein like
monomeric HIV protease (99 residues), this is not an issue since, with an extended-atom model, the storage of
its Hessian takes $\approx$ 10 Mo. 
For dimeric citrate synthase (2$\times$ 450 residues), it takes 3 Go.
Twenty years ago, this used to be an issue~\cite{Marques:95}. It is not any more. 
However, for not so uncommon multimeric systems like aspartate transcarbamylase (nearly 3,000 residues),
using standard approaches would still requires to have access, like at the time~\cite{Perahia:96},
either to smarter numerical methods and/or to supercomputers, 
as the storage of the whole Hessian would take 30 Go of computer memory.

\subsection{Useful methods and approximations}

To overcome this practical limitation, several methods have been proposed.
Note that within the frame of most of them it is not possible to get the $3N$ normal modes of the system,
that is, its full spectrum. 

When a good enough guess of a required eigenvector can be done, for instance when it is expected
to be a relative motion of well-defined structural domains, a method of choice is the Lanczos one,
especially because it only involves calculations of matrix-vector products.
This method was applied to the case of human lysozyme~\cite{Brooks:85} as well as, later on,
using a more sophisticated version of the algorithm, to the case of dimeric citrate synthase~\cite{Marques:95}. 

\begin{figure}[t!]
\includegraphics[width=8.5 truecm,clip]{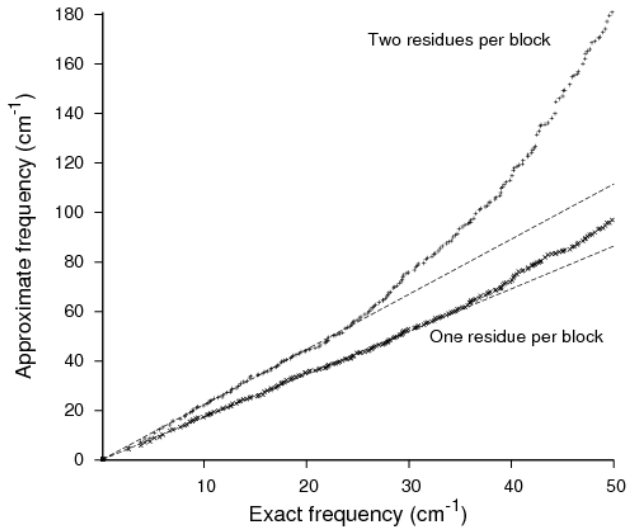}
\caption{\small \label{Fig:rtb}
Approximate frequencies of the HIV protease, as obtained
through the RTB approximation, as a function of exact ones, when one or two amino-acid residues
are put in each block. 
Dotted lines are linear fits for the lowest-frequencies. 
}
\end{figure}

Other methods usually rely on the splitting of the Hessian into blocks~\cite{Harrison:84,Perahia:93},
and/or on the choice of new coordinates that allow for the building of a smaller Hessian
whose eigensolutions are as close as possible to the original ones~\cite{Brooks:12}.
For instance, the principle of the RTB approximation~\cite{Durand:94} is to use
the six rotation (R) and translation (T) vectors of "blocks" (B) of atoms as the new set of coordinates~\footnote{
This approximation has been implemented in CHARMM under the BNM (Block Normal Modes) acronym~\cite{BNM}. It is used by the Eln\'emo Web-server~\cite{Elnemo2}
as well as by softwares like PHASER~\cite{McCoy:Phaser} and DIAGRTB. The later one
is public domain. It can be downloaded at http://ecole.modelisation.free.fr/modes.html.}.
When each block contains a whole amino-acid residue, since there is, on average, $\approx 16$ atoms (48 coordinates) per residue, 
the order of the Hessian is reduced by a factor of $\approx 8$. Interestingly,
if the frequencies calculated with this approximation are found to be, as expected, higher
than those of the full Hessian, on the low-frequency part of the spectrum a proportionality 
is observed between approximate and exact values~\cite{Tama:00} (Fig.~\ref{Fig:rtb}). Moreover, the corresponding
approximate eigenvectors are found to be remarkably similar to exact ones~\cite{Durand:94,Tama:00}.  
Of course, several amino-acid residues can be put in a given block~\cite{Durand:94}, and the way
atoms are grouped into blocks can be made on smarter grounds~\cite{Thorpe:06,Gohlke:06}.

Note that reducing the size of the Hessian is a coarse-graining process, which has the advantage of
preserving the interactions between blocks, as they are observed in the all-atom model. On the other
hand, the fact that low-frequency modes are little perturbed by the process suggest that
they are robust, in the sense that the corresponding pattern of displacements does not depend significantly
upon the details of the description of the system.  

\begin{figure}[t]
\includegraphics[width=8.5 truecm,clip]{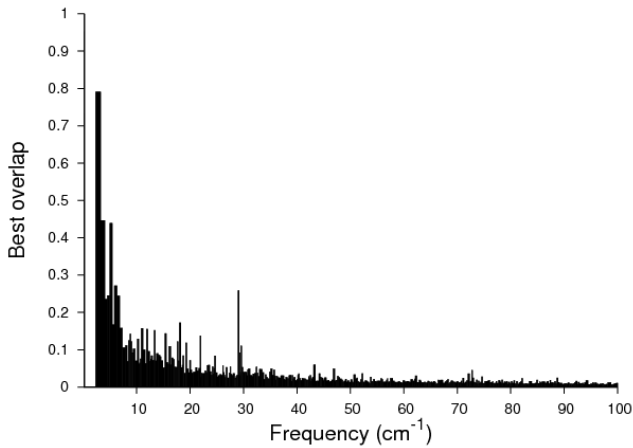}
\caption{\small \label{Fig:bstrob}
For each eigenvector of the HIV protease, as obtained using a standard empirical energy function, 
the best overlap with an eigenvector obtained using Tirion's model is given.
The structure considered is the monomer found in PDB 1HHP, after energy-minimization. 
Tirion's model was built using $R_c = 5${\AA}.
}
\end{figure}

\section{Network Models}

\subsection{Tirion's model}

The fact that an approximation like (\ref{eq:approx}) can prove useful for the study of protein functional
motions is far from being obvious. 
First, because NMA is usually performed for a single minimum
of the potential energy surface, while it is well known that at room temperature a protein explores
a huge number of different ones~\cite{Elber:87}. More generally, (\ref{eq:approx}) means
that the system is studied as if it were a solid, while it is quite clear that the liquid-like character
of the dynamical behaviour of a protein, noteworthy of its amino-acid sidechains~\cite{Kneller:94}, 
is required for its function~\cite{Teeter:01}. 

Moreover, prior to the development of simple implicit solvent models like EEF1~\cite{EEF1}, NMA was usually performed
{\it in vacuo}, while it is well known that water is essential for protein function~\footnote{For its proper folding, in the first place.},
not to mention the significant distortion a system experiences during the required preliminary energy minimization,
when solvent effects are only taken into account through an effective dielectric constant~\cite{Ma:97}.

So, from the very beginning, the usefulness of NMA in the field of protein dynamics
had to mean that functionally relevant normal modes are robust ones, 
similar from a minimum of the energy function to another~\cite{Smith:96,Perahia:10}, but also from a forcefield to another.
As an extreme check of the later point, Monique Tirion proposed to replace electrostatics and Lennard-Jones interactions
in a protein by an harmonic term such that:
\begin{equation}
  \label{eq:tirion}
  V_{nb}=\frac{1}{2} k_{enm} \sum_{d_{ij}^0 < R_c} (d_{ij}-d_{ij}^0)^2
\end{equation}
where $d_{ij}$ is the actual distance between atoms $i$ and $j$, $d_{ij}^0$ being their
distance in the studied structure~\cite{Tirion:96}. In other words, in Tirion's model, all pairs
of atoms less than $R_c$ {\AA}ngstr\"oms away from each other are linked by Hookean springs. 
Note that when, as herein, $k_{enm}$, the force constant, 
is the same for all atom pairs, it only plays the role of a scaling factor for the frequencies of the system.
As a corollary, the only parameter of the model is $R_c$.

\begin{figure}[t]
\includegraphics[width=8.5 truecm,clip]{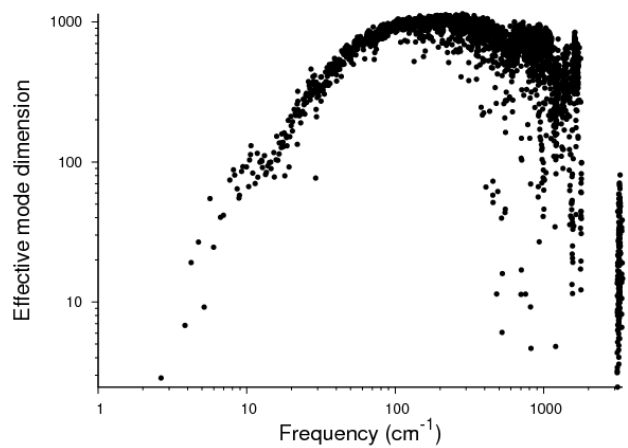}
\caption{\small \label{Fig:neff}
For each eigenvector of the HIV protease, as obtained using a standard empirical energy function, its dimension, \textit{i.e.},
the effective number of Tirion's eigenvectors involved in its description is given.
The structure considered is the monomer found in PDB 1HHP, after energy-minimization. 
Tirion's model was built using $R_c = 5${\AA}.
}
\end{figure}

To evaluate how similar modes obtained with such a description of the non-bonded interactions in a protein
are to those obtained with a standard, empirical, one, a useful quantity is the overlap:
\begin{equation}
O_{ij} = ( \sum_{k}^{3N} a_{ki} a_{kj} )^2
\label{eq:overlap}
\end{equation}
{\it i.e.} the square of the scalar product of the eigenvectors obtained in both cases.
Note that, because eigenvectors are normalized, for each of them: $\sum_{j}^{3N} O_{ij} = 1$.

As shown in Fig.~\ref{Fig:bstrob}, the overlap between an eigenvector obtained with
(\ref{eq:tirion}) and an eigenvector obtained with the energy function available in CHARMM~\cite{EEF1},
can be as high as 0.8, like in the case of the lowest-frequency mode of the HIV protease.
So, obviously, this mode is a very robust one, in the sense that it depends little upon the nature 
of the atomic interactions in the structure. Note in particular that, 
in the simplified version of Tirion's model considered herein, 
at variance with Tirion's original work~\cite{Tirion:96},
chemically bonded atoms are
treated on the same footing as other pairs of close atoms, that is, by linking them with 
a spring of same force constant ($k_{enm}$).

\begin{figure}[t]
\includegraphics[width=8.5 truecm,clip]{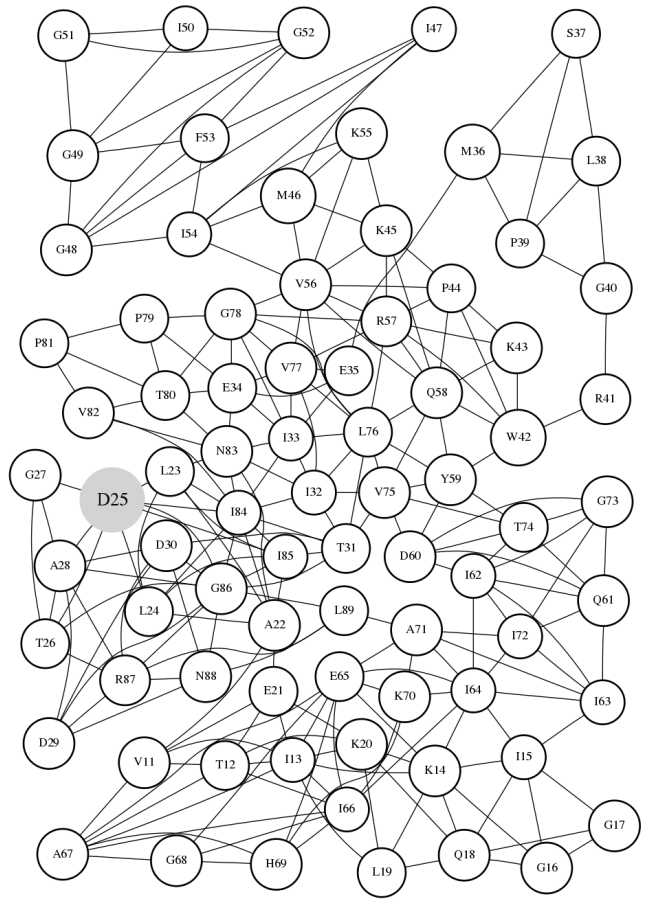}
\caption{\small \label{Fig:graph}
Graph of the interactions between HIV-protease residues.
The catalytic residue (Asp 25) is indicated in grey.
For the sake of clarity, the ten first and last residues were omitted. 
Drawn with Graphviz-Neato.
}
\end{figure}

In the HIV protease case, only three 
low-frequency modes have an overlap with a mode obtained using Tirion's model that is larger than 0.4.
However, this does not necessary mean that HIV protease has only three highly robust modes, 
in particular because a given subset of modes can prove robust as a whole, even when each of them is not.  
To quantify this, $D_i$, the dimension of standard mode $i$, {\it i.e.},
the effective number of Tirion's modes involved in its description, can
be calculated as follows~\cite{Bruschweiler:95,Tama:01}:
\begin{equation}
D_i = exp( - \sum_j^{3N} O_{ij} log O_{ij} )
\label{eq:neff}
\end{equation}
For the lowest-frequency mode of the HIV protease, $D_1 =$ 2.8 and, indeed, as shown in Fig.~\ref{Fig:neff},
on the low-frequency side of the spectrum only
three standard modes of the HIV protease can be described accurately with less than 10 Tirion's modes.
However, three other ones can be described with less than 30 of them. Note also that some high-frequency modes can 
be described with a handful of Tirion's modes. This is because atomic masses are taken into account
in Tirion's model. As a consequence, within the frame of this model also, 
high-frequency modes are well localized and, likewise, they correspond to localized motions 
in which hydrogen atoms are involved.

\begin{figure}[t]
\includegraphics[width=8.5 truecm,clip]{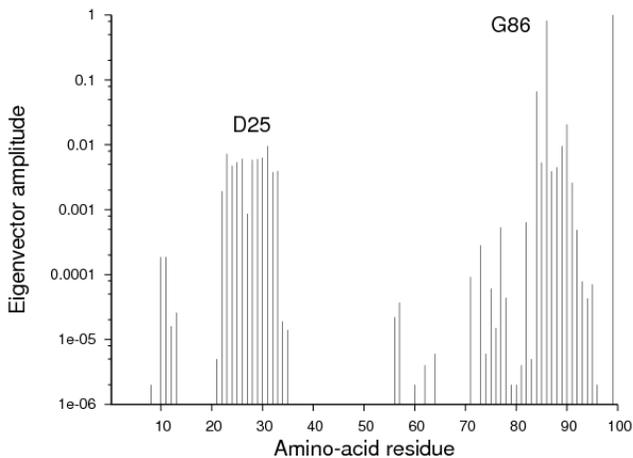}
\caption{\small \label{Fig:gnm}
Eigenvector of the adjacency matrix of the HIV-protease (PDB 1HHP), corresponding to the largest eigenvalue. 
Here, two residues are defined as interacting ones ({\it i.e.}, the corresponding matrix element is -1)
when the distance between their C$_\alpha$ is less than 6.5{\AA}. Note the logarithmic scale.
}
\end{figure}

\subsection{Proteins as undirected graphs}
\label{sec:graph}

Describing interactions between amino-acid residues in terms of short-range harmonic springs (with, \textit{e.g.}, $R_c$=5{\AA}), 
as proposed by Monique Tirion, and ending with a subset of low-frequency modes that are highly similar 
to those obtained with a much more complex description, suggests that such modes   
are due to some generic property. 

A possibility is that what determines the nature of these modes is the pattern of interactions
between residues. As a matter of fact, simple methods have been proposed for evaluating
protein flexibility through the study of the graph corresponding to the set of interactions
observed in a given structure (Fig.~\ref{Fig:graph} shows such a graph, in the case of the HIV protease).  
For instance, it has been shown that, using the highly efficient pebble game algorithm~\cite{Thorpe:95},
hinges and flexible loops can be identified in proteins like the HIV protease, 
adenylate kinase, \textit{etc}~\cite{Thorpe:01}.
On the other hand, diagonalizing directly the corresponding, so-called adjacency, matrix provides 
fair estimates of the amplitude of atomic fluctuations, as observed experimentally,
noteworthy through crystallographic Debye-W\"aller factors~\cite{Erman:97,Phillips:06} (see Section~\ref{sec:bf}). 

Interestingly, while the small eigenvalues of the adjacency matrix
are enough for providing such estimates, large eigenvalues seem also able to
provide useful informations. Indeed, 
because eigenvectors corresponding to large eigenvalues pinpoint 
spots in the structure where residue density (in a coarse-grained sense) is the highest,
it has been suggested that they may correspond to protein folding cores~\cite{Erman:98}.

For instance, in the case of the HIV protease~\cite{Erman:98},
the eigenvector corresponding to the largest eigenvalue is dominated by
the motion of Gly 86 (Fig.~\ref{Fig:gnm}) a residue
close to Asp 25, the catalytic residue (see Fig.~\ref{Fig:graph}).
As a matter of fact, there is an hydrogen-bond between the backbone carbonyl 
oxygen of the previous residue, Ile 85, and the backbone amide nitrogen of Asp 25.
The fact that, in such a small protein, the folding core could prove that close
to the enzyme active site would make sense, since enzymatic activity usually requires
a precise positioning of the residues involved in the catalytic mechanism, that
is, a rather rigid local environment.

\begin{figure}[b]
\includegraphics[width=8.5 truecm,clip]{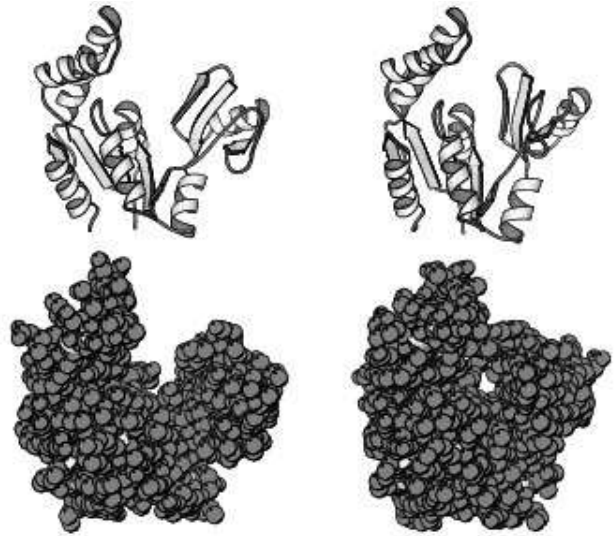}
\vskip -4.0 mm
\caption{\small \label{Fig:dconfgk}
Conformational change of {\it Saccharomyces cerevisiae} guanylate kinase.
Left: open form (PDB 1EX6). Right: closed form (PDB 1EX7). Top: standard sketch.
Bottom: van-der-Waals spheres. Drawn with Molscript~\cite{Molscript}.
}
\end{figure}

\subsection{Elastic Networks}

However, by considering a protein as a graph, an important information is lost, namely,
the directionality of the motions. This is probably why, among
the family of simple protein models, 3D-Elastic Network models (ENM) 
seem to be the most popular ones. Such models
are just one simplification step further from Tirion's model: here, residues are
described as beads, often a single one, as initially proposed~\cite{Hinsen:98},
although beads can also represent groups of residues~\cite{Bahar:05}, as well as whole protein
monomers~\cite{Tama:02v}. 

Using as beads the C$_\alpha$ atoms (C$_\alpha$-ENM), it was shown that a few low-frequency modes
of the coarse-grained elastic networks thus obtained are enough for describing accurately
the motion a protein experiences upon ligand binding~\cite{Tama:01,Delarue:02,Gerstein:02},
as long as a significant portion of
the protein is involved~\cite{Tama:01} (\textit{e.g.}, whole domains), 
at least when the amplitude of the motion is large enough~\cite{Nicolay:06}
(typically, more than $\approx$ 2{\AA} of C$_\alpha$-rmsd~\footnote{rmsd: root-mean-square deviation.}).

\begin{figure}[t]
\includegraphics[width=8.5 truecm,clip]{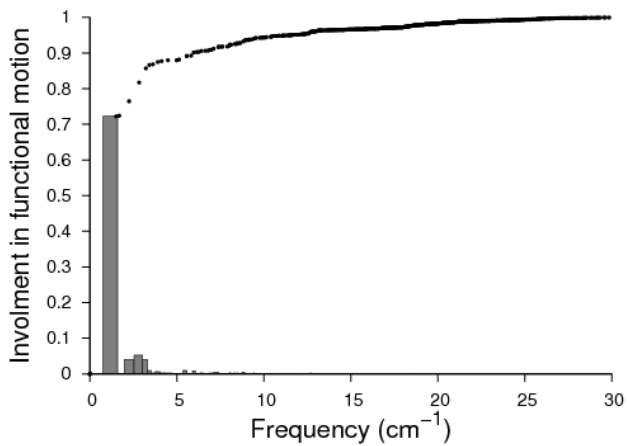}
\caption{\small \label{Fig:overlapgk}
Accuracy of the description of the
conformational change of {\it Saccharomyces cerevisiae} guanylate kinase with low-frequency modes,
as obtained using an elastic network model (PDB 1EX6, $R_c$ = 10 {\AA}).
Boxes: involvement coefficients. Black disks: cumulative sum.
The force constant of the springs was chosen so as to have a lowest-frequency of 1.5 cm$^{-1}$.
}
\end{figure}

To measure how well normal mode $i$ describes a given motion, a useful quantity is $I_i$,
its involvement coefficient~\cite{Marques:95,Ma:97}:
\begin{equation}
I_{i} = ( \sum_{j}^{3N} a_{ji} \frac{\Delta r_j}{| \vec{\Delta r} |} )^2
\label{eq:involv}
\end{equation}
where $\vec{\Delta r} = \vec{r_b} - \vec{r_a}$, $\vec{r_a}$ and $\vec{r_b}$
being the atomic positions observed in conformations $a$ and $b$, respectively.
Note that in order to have meaningful involvement coefficients, conformation $b$
needs to be fitted onto conformation $a$, if the normal modes were obtained for the later. 
Note also that (\ref{eq:involv}) is the square of the scalar product between $\vec{\Delta r}$ and
normal mode $i$. So, as a consequence of normal modes orthogonality: $\sum_i^{3N} I_{i} = 1$.  
In other words, $I_i$ gives the fraction of the protein motion that can be described just by
considering the displacement of the system along mode $i$.

\begin{figure}[t]
\includegraphics[width=8.5 truecm,clip]{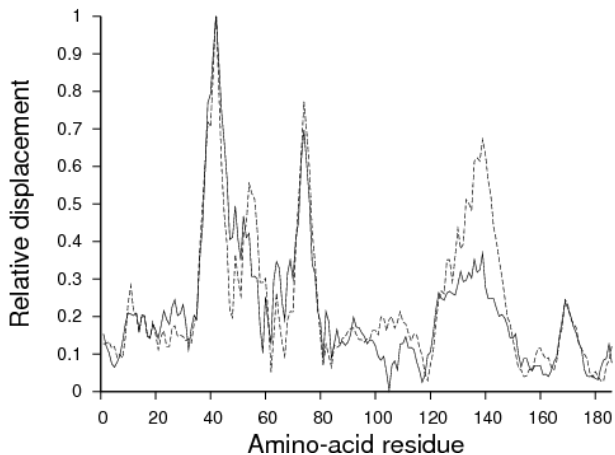}
\caption{\small \label{Fig:gkmod7}
Comparison of the 
conformational change of {\it Saccharomyces cerevisiae} guanylate kinase (plain line) with its lowest-frequency mode (dotted line),
as obtained using an elastic network model (PDB 1EX6, $R_c$ = 10 {\AA}).
Both sets of C$_\alpha$ displacements are normalized with respect to the maximum value.
}
\end{figure}

$\sum_i^{n} I_{i}$,
the quality of the description of the closure motion of guanylate kinase (Fig.~\ref{Fig:dconfgk}),
as a function of $n$, the number of modes taken into account, is shown in Fig.~\ref{Fig:overlapgk}, for the modes
of the C$_\alpha$-ENM built with an open form (R$_c$ = 10{\AA}). In this case,
which is far from being an exceptional one~\cite{Tama:01,Gerstein:02}, the lowest-frequency mode
is enough for describing 72\% of the functional motion ($I_1$ = 0.72). 
Note that such a high value of the involvement coefficient means that both patterns of atomic displacements are remarkably similar (Fig.~\ref{Fig:gkmod7}).
Indeed, the main difference concerns the relative amplitude of the motion of helices 
125-135 and 141-157.
On the other hand, together, modes three to five are able to describe 13\% of the motion ($I_3$ = 0.04, $I_4$ = 0.05, $I_5$ = 0.04).
So, four modes {\it only} are enough for describing 85\% of the  closure motion of guanylate kinase (\ref{Fig:overlapgk}).

\subsection{Robust modes}

However, to turn such a qualitative description into a possibly useful prediction,
a way to identify {\it a priori} the modes the most involved in the functional motion is required. 
To do so, low-resolution experimental data can prove enough like, for instance,
those obtained by cryo-electromicroscopy~\cite{Tama:03,Delarue:04,NORMA}.
A more general possibility is to build upon the robustness of these modes,
namely, to look for modes that are little sensitive to the protein model used~\cite{Nicolay:06}.
For instance, instead of setting springs between pairs of $C_\alpha$ atoms less than $R_c$ {\AA}ngstr\"oms away
from each other, as above, the springs can be established so that each $C_\alpha$ atom is linked
to $\approx n_c$ of its closest neighbors~\cite{Nicolay:06}.  
Then, overlaps between both sets of modes can be obtained (Eq.~\ref{eq:overlap}) and when a mode
can be described accurately with a small number of modes (Eq.~\ref{eq:neff}) of the other set 
it is considered to be a robust one. In the case of the open form of guanylate kinase, with $R_c$ = 10 {\AA}
and $n_c$ = 10, the first four modes calculated with $R_c$ = 10 {\AA} can be described
with less than three modes calculated with $n_c$ = 10, while all others need more than eight. 
Note that the four robust modes thus identified are enough for describing 81\% of the closure motion of guanylate kinase
(mode two does not contribute at all; see Fig.~\ref{Fig:overlapgk}).

\begin{figure}[t]
\includegraphics[width=8.5 truecm,clip]{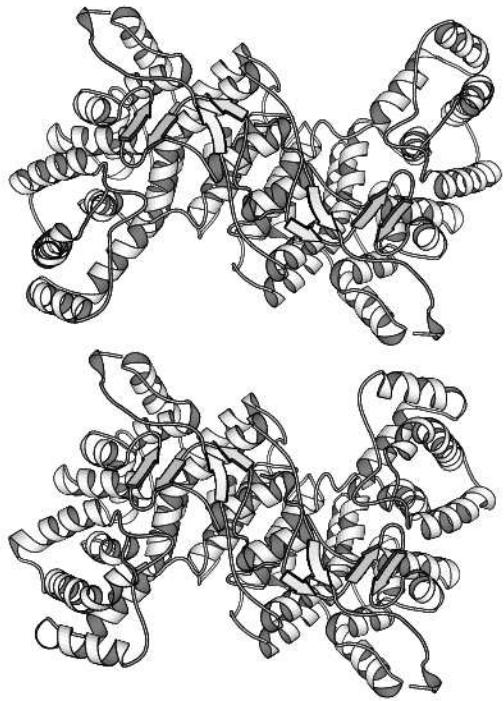}
\vskip -4.0 mm
\caption{\small \label{Fig:dconfcs}
Conformational change of {\it Thermus thermophilus} citrate synthase.
Top: open form (PDB 1IOM). Bottom: closed form (PDB 1IXE). Drawn with Molscript~\cite{Molscript}.
}
\end{figure}

\begin{figure}[t]
\includegraphics[width=8.5 truecm,clip]{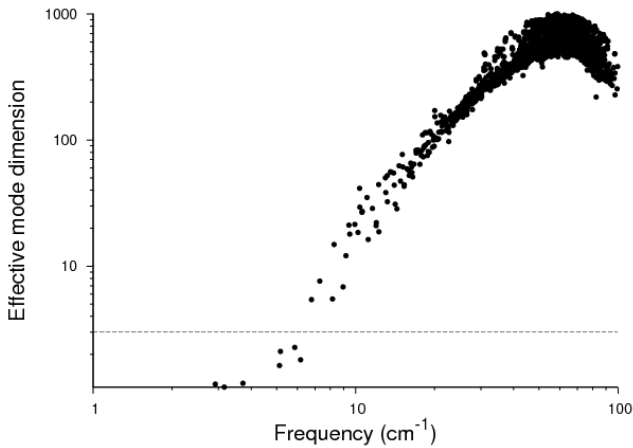}
\caption{\small \label{Fig:csneff}
Effective dimension of each eigenvector of dimeric citrate synthase, as obtained with a C$_\alpha$-ENM (PDB 1IOM; $R_c$ = 10).
Here, the effective dimension of a mode corresponds to
the number of eigenvectors involved in its description, when the later are obtained with $n_c$ = 10.
The force constant of the springs was chosen so as to have a lowest-frequency of 2.9 cm$^{-1}$, with $R_c$ = 10.
The dotted line indicates an effective dimension of three.
}
\end{figure}

The case of guanylate kinase may look too simple, since the functional motion
can be correctly guessed just by looking at the structures (see Fig.~\ref{Fig:dconfgk}).
Interestingly, results obtained with this model system seem to have a general character~\cite{Nicolay:06}.
For instance, in the case of citrate synthase, it is possible to guess where the active site is and, 
as a consequence, where the closure motion should occur. However, the
structure is more complex (Fig.~\ref{Fig:dconfcs}) and it is hardly feasible to decide where are
the limits of each structural domain. 
Nevertheless, results obtained through a normal mode analysis of a C$_\alpha$-ENM of dimeric citrate synthase
are almost as impressive as those obtained for guanylate kinase.
Noteworthy, with $R_c$ = 10 {\AA}, $I_2$ = 0.29 and $I_3$ = 0.48, which means that
77\% of the conformational change of citrate synthase can be described with these two modes.
On the other hand, with $n_c$ = 10, $I_2$ = 0.27 and $I_3$ = 0.48. 
As expected, modes two and three are among the most robust ones. However, 
in the case of dimeric citrate synthase, there are more than two robust modes. Actually, the first seven modes obtained with $R_c$ = 10
can be described accurately with less than three modes obtained with $n_c$ = 10 (Fig.~\ref{Fig:csneff}).

Note that, when compared with modes obtained with $n_c$ = 10, high-frequency modes obtained with $R_c$ = 10
can certainly not be considered as being robust (their effective dimension is 200 or more; Fig.~\ref{Fig:csneff}).
This is because, as a consequence of the cutoff criterion, such modes correspond to motions of residues belonging to parts of the structure where 
density (in a coarse-grained sense) is the highest. This is not the case with $n_c$ = 10. As a matter of fact,
the later kind of ENM was designed so as to show that low-frequency modes do not result from density patterns
inside a structure but from mass distribution in space, that is, from the overall shape of the structure~\cite{Nicolay:06}.

\begin{figure}[t!]
\includegraphics[width=8.5 truecm,clip]{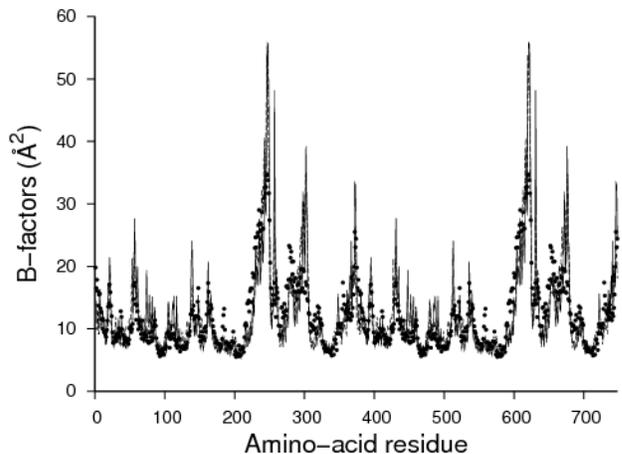}
\caption{\small \label{Fig:bfcit}
B-factors of dimeric citrate synthase.
Filled circles: experimental values (PDB 1IOM).
Plain and dotted lines, respectively:
values calculated either with a standard ENM ($R_c$ = 10) or with $n_c$ = 10.
Calculated values are shifted and scaled, so that minimum and average values are the same as
experimental ones. A justification for such a treatment is that experimental B-factors
include various physical effects, other than thermal intra-protein fluctuations. 
}
\end{figure}

\subsection{Crystallographic B-factors}
\label{sec:bf}

Having access to an analytical solution of the equations of atomic motion (Eq.~\ref{eq:fluct})
allows for the calculation of many quantities~\cite{NMA}, like the fluctuations of atomic coordinates around
their equilibrium values. For instance, in the case of the $x$ coordinate of atom $i$:
\begin{equation}
<\Delta x_i^2> = k_B T \sum_{k=1}^{n} \frac{a_{ik}^2}{4 \pi^2 m_i \nu_k^2}
\label{eq:bfluct}
\end{equation}
where $T$ is the temperature, $k_B$, the Boltzman constant, $n$, the number of modes taken into account
($n = 3N-6$, unless specified otherwise)
and where $<>$ denotes a time average.
Note that the six rigid-body modes (translation and rotation ones), as a consequence of
their null frequencies, are usually excluded 
from (\ref{eq:bfluct}). 

Interestingly, crystallographic B-factors are expected to derive from the fluctuations of atomic
positions within the crystal cell, namely:
\[
B_i = \frac{8 \pi^2}{3} <\Delta x_i^2 + \Delta y_i^2 + \Delta z_i^2>
\]
where $B_i$ is the isotropic B-factor of atom $i$.
Although other factors contribute to the experimental values, such as crystal disorder or phonons,
fair correlations have been obtained with values calculated using ENMs~\cite{Bahar:97}, especially 
when effects of neighboring molecules in the crystal are included~\cite{Phillips:02,Hinsen:08}.
Fig.~\ref{Fig:bfcit} illustrates how accurate B-factors calculated with ENMs can be, even without taking such subtleties into account.
Here, the correlation between experimental and calculated values is 0.80, when
a cutoff-based ENM is used, and 0.85, when a constant-number-of-neighbors ENM is used.
Note that predictions made with both kinds of ENMs are hardly distinguishable.
This is a mere consequence of the weight of the low-frequency modes in (\ref{eq:bfluct}).
Indeed, taking into account only the seven robust modes identified previously ($n=7$)  
yields a correlation of 0.82, that is, a value as high as when all modes are used ($n=3N-6$).

\subsection{Flexibility versus rigidity}

As briefly discussed in Section \ref{sec:graph}, while low-frequency modes 
can provide information on the overall flexibility of a structure, 
when a cutoff-based C$_\alpha$-ENM is considered, high-frequency modes 
pinpoint parts of the structure where amino-acid density is the highest.
Though it is tempting to view such parts as being the most rigid ones,
the relationship between protein rigidity and density is not expected to be that
straightforward. Also, because high-frequency modes are usually localized,
several of them need to be considered in order to get a consistent picture of the overall rigidity
inside a structure. Unfortunately, selecting the corresponding subset of high-frequency modes in a rigorous way
is not that obvious.

\begin{figure}[t]
\includegraphics[width=8.5 truecm,clip]{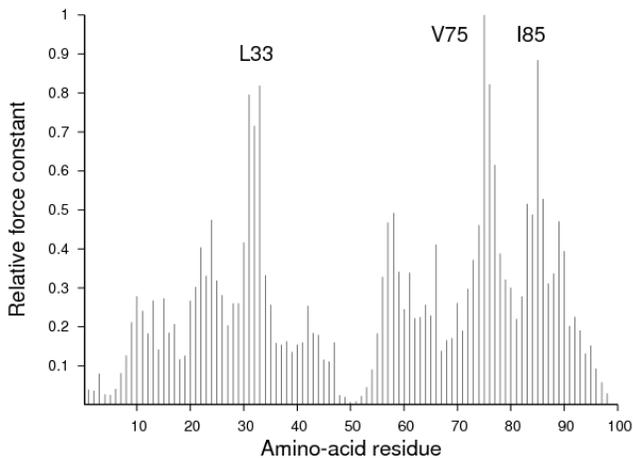}
\caption{\small \label{Fig:hivrig}
Rigidity of a monomer of the HIV-protease (PDB 1HHP, $R_c$ = 10).
Effective force constants have been normalized so that the maximum value is one.
}
\end{figure}

Several alternatives have been proposed. 
For instance, $k_i$, an effective local force constant associated to atom $i$, can be defined as follows~\cite{Lavery:06,Lavery:07}:
\[
k_i = \frac{3 k_B T}{<(\bar{d_i} - <\bar{d_i}>)^2>}
\]
where $\bar{d_i}$ is the average distance of atom $i$ from all other atoms in the structure.
Note that it is when this average distance fluctuates little that the force constant (the rigidity) is high.
Of course, when the considered protein is a multi-domains one,
the average is meaningful only if it is calculated for atoms belonging to the same domain.

Interestingly, this measure involves an ensemble averaging that can be performed
using any protein model, \textit{e.g.}, all-atom as well as coarse-grained ones.
For instance, a C$_\alpha$-ENM can be used together with Eq.~\ref{eq:fluct}. 
Fig.~\ref{Fig:hivrig} shows the result in the case of the HIV protease.
Like with the top eigenvector of the adjacency matrix (see Fig.~\ref{Fig:gnm}),
the peptidic stretch nearby G86 is identified as being rigid, 
but the site now identified as being the most rigid, V75, was not pinpointed by the previous approach. 
However, V75 is found to be significantly involved in the eigenvector corresponding to the fourth highest eigenvalue
of the adjacency matrix.
This suggests that when several eigenvectors are taken into account, 
both methods may provide similar informations about the rigidity of protein structures.     

\begin{figure}[t]
\includegraphics[width=8.5 truecm,clip]{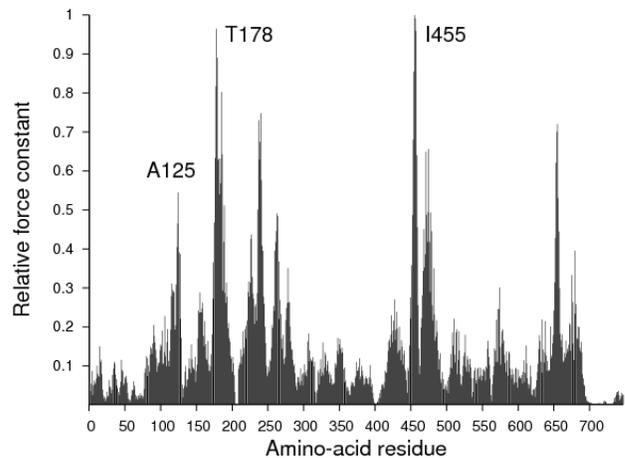}
\caption{\small \label{Fig:myorig}
Rigidity of \textit{Dictyostelium discoideum} myosin (PDB 1VOM, $R_c$ = 10).
}
\end{figure}

Fig.~\ref{Fig:myorig} shows what such an analysis yields in the case of a larger protein,
namely, myosin (747 amino-acid residues). Interestingly, T178 and I455, the sites
identified as being the most rigid, are both quite close to the enzyme active site,
their carbonyl oxygens being 6 and 4{\AA} away, respectively, from phosphate oxygens belonging
to the ATP analogue observed in the studied structure (PDB 1VOM). 

\subsection{Non-linear network models}

Elastic network models have proved useful, in particular because they are very simple.
As a consequence, analyses performed with such models can be very quick,
allowing for an almost instantaneous check of an hypothesis as well as for
large scale studies. 
However, for many specific applications, ENMs are expected to prove too naive.
Therefore, it is of interest to develop more complex models. 
But, because when complexity increases, the time it takes to perform an analysis usually also does (either for a human being or for a computer),
the most useful are expected to be models that are only a bit more complex than ENMs.
In practice, since ENMs are, in essence, single-parameter models,
increasing their complexity means increasing their number of parameters.
So, in order not to increase their complexity too much,
a natural choice is to keep
the number of parameters as small as possible. 

Along this line of thought, 
in order to recover one of the major property of the potential energy surface of a protein, namely, that it is 
a multi-minima one, network models with several (usually two) energy minima have been proposed~\cite{Karplus:05,Voth:07}.
However, a more basic property of such a surface is that it is highly anharmonic~\cite{Levy:82,Hayward:95}.
So, it seems worth starting by adding explicit non-linearity~\footnote{Anharmonicity and non-linearity may sound more familiar to, respectively, biophysicists and physicists.}
into an elastic network model, \textit{e.g.}~\cite{Juanico:07}:
\begin{equation}
  \label{eq:nnm}
  V=\sum_{d_{ij}^0 < R_c} \frac{k_2}{2} (d_{ij}-d_{ij}^0)^2 + \frac{k_4}{4} (d_{ij}-d_{ij}^0)^4
\end{equation}
the choice of an additional term with a power of four, instead of three, being mostly for symmetry reasons
but also because previous works on one and two-dimensional systems had shown that the dynamical properties
of systems with such an energy function can be quite spectacular~\cite{Flach:94,DB:04}.
Note that when $k_4 = 0$, (\ref{eq:nnm}) corresponds to (\ref{eq:tirion}), that is, to a standard
ENM (up to now, only protein models with $\frac{k_4}{k_2} =$ 1 {\AA}$^2$ have been studied in depth~\cite{Piazza:11}). 

One of the dynamical non-linear phenomenons that can occur in this context is the birth and the (rather) long-time survival of
a discrete breather (DB)~\cite{Flach:94}, that is, a localized mode whose frequency is high enough, so that energy exchange
with the rest of the system can prove extremely slow (thousands of periods of the DB). A way   
to observe such a phenomenon without making any a priori assumption on its nature (localized or not, \textit{etc}) is to perform
a molecular dynamics simulation, starting with a high initial temperature and cooling the system through
friction on its surface atoms~\cite{Juanico:07}. When a DB sets up, the energy remaining into the system becomes
(quite suddenly) localized, the few atoms involved in the DB being the only ones with a significant motion
whereas all other ones are almost frozen. 

\begin{figure}[t]
\includegraphics[width=8.5 truecm,clip]{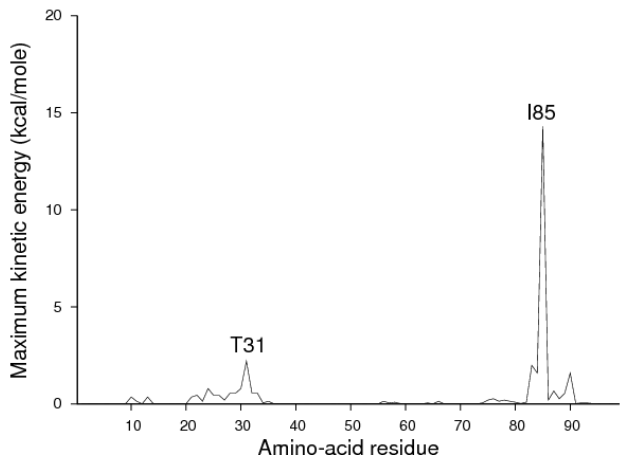}
\caption{\small \label{Fig:dbhiv}
Discrete breather arising from the highest-frequency mode of the HIV-protease (PDB 1HHP, $R_c$ = 10).
Maximum kinetic energies observed during a 200 ps MD simulation are given, when
the highest-frequency mode of the elastic network is excited with 20 kcal/mole.
}
\end{figure}

Interestingly, DBs tend to appear in the most rigid parts of a structure~\cite{Juanico:07}.
This is due to the fact that most of them are related to one of the high-frequency modes of the elastic network. 
As a matter of fact, the more energetic a DB is, 
the higher its frequency~\cite{Juanico:07}, the more localized~\cite{Juanico:07,Piazza:08} and the
more different it is from the high-frequency mode it comes from~\cite{Piazza:08}.
As a consequence, a way to obtain DBs is to provide energy to one of the high-frequency 
modes of the network~\cite{Sanejouand:09}. Because energetic DBs are highly localized, such a protocol allows
to pinpoint a few specific residues (a single one per high-frequency mode~\footnote{If the energy provided is high enough.}).

\begin{figure}[t]
\includegraphics[width=8.5 truecm,clip]{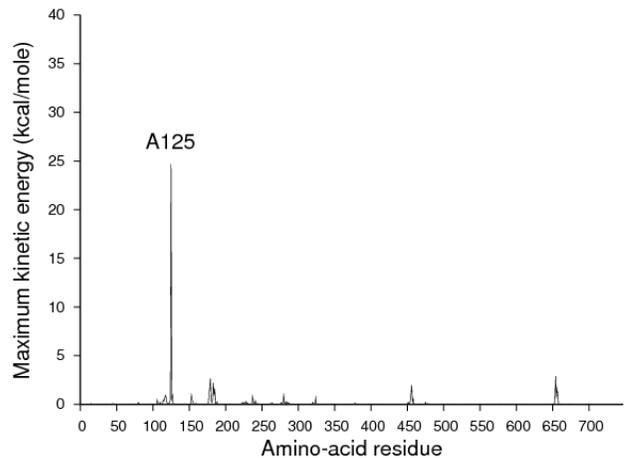}
\caption{\small \label{Fig:dbmyo}
Discrete breather arising from the highest-frequency mode of \textit{Dictyostelium discoideum} myosin (PDB 1VOM, $R_c$ = 10).
Maximum kinetic energies observed during a 200 ps MD simulation are given, when
the highest-frequency mode of the elastic network is excited with 40 kcal/mole.
}
\end{figure}

For instance, in the case of the HIV-protease, exciting the highest-frequency mode 
highlights residues Ile 85 and Thr 31 (compare Fig.~\ref{Fig:dbhiv} to Figs.~\ref{Fig:gnm} and~\ref{Fig:hivrig})
while, in the case of myosin, such an approach highlights residue Ala 125 (see Fig.~\ref{Fig:dbmyo} and compare to Fig.~\ref{Fig:myorig}).
Note that in both cases, more than 50\% of the kinetic energy initially given to the highest-frequency mode of the system
is observed, during the simulation, on a single residue~\footnote{The lifetime of these discrete breathers is longer than 200 ps.}. 

\section{Conclusion}

The study of simple models where a protein is described as a set of Hookean springs 
linking neighboring amino-acid residues often provides useful informations about its flexibility.
Noteworthy, the correlation between calculated residue fluctuations and experimental B-factors 
can be quite high (Fig.~\ref{Fig:bfcit}). Moreover, in many cases, a few low-frequency modes 
are found able to provide a fair description of the functional motion of a protein (see Fig.~\ref{Fig:overlapgk} and~\ref{Fig:gkmod7}).
This is far from being an obvious result since, for instance, the energy function of a set of Hookean springs
has a single minimum, while a conformational change is expected to involve (at least) two significantly different minima
of the potential energy surface. 

Interestingly, modes involved in functional motions were also found to be robust~\cite{Nicolay:06}, that is, very
little sensitive to changes in the model used to describe the protein. Actually, the robustness of a small
subset of the lowest-frequency modes of a protein explains why coarse-grained models can be used on the same footing as
highly detailed ones, as far as low-frequency (and large amplitude) motions are concerned.
Reciprocally, seeking for robust modes allows to get a small set of coordinates (eigenvectors)
able to provide a fair description of the functional motion a given protein can perform.

More surprisingly, 
high-frequency modes of cutoff-based elastic networks may also prove useful.
Indeed, they pinpoint parts of the structure where the residue density (in a coarse-grained sense)
is the highest, and it seems that such parts can be important for the proper folding~\cite{Erman:98} and/or
for the function~\cite{Bahar:05,Lavery:07,Juanico:07} of many proteins.


\end{document}